\documentclass{comnet}%%%%where comnet is the template name
\begin{document}

\title{Controllability of a Class of Swarm Signaling Networks}

\shorttitle{Controllability of a Class of Swarm Signaling Networks} %%%for recto running head
\shortauthorlist{P. Sun \emph{et al}} %%% for verso running head

\author{%%%% First author details
\name{Peng Sun}
\address{Delft University of Technology, Delft, The Netherlands}
\name{Robert E. Kooij$^*$}
\address{Delft University of Technology, Delft, The Netherlands\\
TNO, Unit ICT, The Netherlands
\email{$^*$Corresponding author: r.e.kooij@tudelft.nl}}
%%%%%%% Second author details
\name{Roland Bouffanais}
\address{University of Ottawa, Ottawa, Canada}
%%%%%%%
}

\maketitle

\begin{abstract}
{In this paper, we propose closed-form analytical expressions to determine the minimum number of driver nodes that is needed to control a specific class of networks. We consider swarm signaling networks with regular out-degree distribution where a fraction $p$ of the links is unavailable. We further apply our method to networks with bi-modal out-degree distributions. Our approximations are validated through intensive simulations. Results show that our approximations have high accuracy when compared with simulation results for both types of out-degree distribution.}
{network controllability; swarm signalling networks; driver nodes}
%%%% If classification number provided then

\end{abstract}

\section{Introduction}
Network controllability is an essential property for the safe and reliable operation of real-world infrastructures, and as such this research area has attracted significant attention over the past decade~\cite{liu2011controllability,yuan2013exact,jia2013emergence,nepusz2012controlling}. For definiteness, a system is said to be controllable if it can be driven from any initial state to any desired final state by external inputs in finite time~\cite{lombardi2007controllability}. By blending classical control theory with concepts from network science, the notion of structural controllability has emerged~\cite{lin}. Classically, let $\mathbf{A}$ be the $N \times N$ adjacency matrix of a given network with $N$ nodes, while the connection of $\mathbf{M}$ input signals to the network is described by the $N \times M$ input matrix $\mathbf{B}$, where $M \leq N$. Then, the system characterized by $(\mathbf{A},\mathbf{B})$ is structurally controllable if it is possible to find the non-zero parameters in $\mathbf{A}$ and $\mathbf{B}$ such that the obtained system $(\mathbf{A},\mathbf{B})$ is controllable in the classical sense of satisfying the Kalman rank condition.

In their seminal article, Liu \emph{et al.}~\cite{liu2011controllability} used maximum matching to get the minimum number $N_D$ of driver nodes---i.e., nodes driven by external inputs---that are needed to achieve structural controllability of a directed network. However, the results reported in Liu \emph{et al.}~\cite{liu2011controllability} critically depend on the assumption that the network has no self-links, i.e. a node's internal state can only be changed upon interaction with neighboring nodes~\cite{cowan}. Yuan \emph{et al.}~\cite{yuan2013exact} further proposed the concept of exact controllability based on the maximum multiplicity of all eigenvalues of the adjacency matrix $A$ to find the driver nodes in networks. Ruths \emph{et al.}~\cite{ruths2014control} developed a theoretical framework for characterizing control profiles of networks. Jia \emph{et al.}~\cite{jia2013emergence} classified each node into one of three categories, based on its likelihood of being included in a minimum set of driver nodes and discovered bi-modal behavior for the fraction of redundant nodes when the average degree of the networks is high. Yan \emph{et al.}~\cite{yan2015spectrum} investigated the relation between the maximum energy needed for controllability and the number of driver nodes. Nepusz \emph{et al.}~\cite{nepusz2012controlling} indicated that most real-world networks are more controllable than their randomized counterparts. More recently, Zhang \emph{et al.}~\cite{zhang2019evolution} studied the change of network controllability in growing networks, and found a lower bound for the maximum number of nodes that can be added to a network while keeping the number of driver nodes unchanged.

The robustness of network controllability under perturbation of the network topology has been investigated extensively. Lu \emph{et al.}~\cite{lu2016attack} discovered that a betweenness-based strategy is quite efficient to harm the controllability of real-world networks. Lou \emph{et al.}~\cite{lou2020towards} present a search for the network configuration with optimal robustness of controllability against random node-removal attacks. Wang~\emph{et al.} \cite{wang2020controllability} proposed a dynamic cascading failure model and investigated the controllability robustness of real-world logistic networks. Nie \emph{et al.}~\cite{nie2014robustness} found that the controllability of Erd\H{o}s-R\'{e}nyi random networks with a moderate average degree is less robust, whereas a scale-free network with moderate power-law exponent shows a stronger ability to maintain its controllability when these networks are under intentional link attack. Sun \emph{et al.}~\cite{sun2019quantifying} proposed closed-form analytic approximations for the minimum number of driver nodes needed to fully control networks, where links are removed according to both random and targeted attacks.
Komareji \emph{et al.}~\cite{Komareji} discussed the resilience and controllability of dynamic collective behaviors for a class of Swarm Signaling Networks (SSNs)~\cite{Komareji}. The SSNs are modeled as directed (unweighted) graphs where the nodes have $k$-regular out-degree and Poisson-like in-degree distribution with average $k$. Following the paper by Liu et al.~\cite{liu2011controllability}, an implicit equation is derived, whose solution leads to the minimum number of driver nodes to control the whole swarm~\cite{Komareji}. However, upon validation of the formula given in~\cite{Komareji} through simulation, we found significant differences between the analytical results and simulation results.

Beyond the theoretical interest in analytical results related to the controllability of complex networks, it is worth stressing that our particular focus on SSNs stems from their practical importance and ubiquity in a number of key problems related to collective behaviors and space-dependent collective decision-making~\cite{Bouffanais2016}. By construction, the nodes of SSNs are embedded in the physical space and the specific nature of inter-agent interactions governs the distribution of edges. Hence, the SSN topology---with its particular in- and out-degree distributions, and high clustering---is a powerful abstraction to study the dynamics of these collective behaviors. For instance, when considering natural swarms---e.g., schools of fish or flocks of birds---the concept of controllability of the SSN is key to explain how a single agent detecting a predator is capable of triggering a collective evasive maneuver~\cite{Komareji,komareji_2014}. The analysis of the controllability of SSNs is even more important when considering artificial swarming systems: e.g., groups of robots collectively moving in space~\cite{sekunda2016}, or performing a decentralized mapping of an open space~\cite{kit2019decentralized}, or aiming at achieving a spatial consensus~\cite{mateo2019}. In all these multi-robot systems, the tuning of the topology of the SSNs plays a key role in achieving the desired collective actions. Even for problems of social contagion in collective decision-making, the Kirchhoff index and clustering coefficients of the SSN have been found to be responsible for a transition from a simple social contagion to a complex one~\cite{horsevad2022transition}. In all these natural, artificial and social systems, the effectiveness in achieving an effective collective response rests on the amplified influence of a few agents (i.e., nodes of the SSN) over the entire network. Therefore, a detailed understanding of the controllability of various types of SSN would offer valuable insights into the complex dynamics of this broad class of collective behaviors.

The aim of this paper is threefold. First, we correct the assumption when calculating the minimum fraction of driver nodes given in \cite{Komareji} and back this up with simulations.
Second, we generalize the results by considering SSNs in which a fraction $p$ of the links are removed at random. Also for this case, we are capable of deriving an implicit equation, whose solution leads to the minimum number of driver nodes. 
Finally, we relax the condition that the out-degree is regular. Specifically, we consider bi-modal out-degree distributions, where the out-degree is $k_1$ for a fraction $\alpha$ of the nodes and $k_2$ for the remaining fraction $(1-\alpha)$ of the nodes. Note that the impact of having unavailable links is also considered a more general scenario.

\section{Controllability of networks and driver nodes}

\subsection{Controllability of networks}
A system is controllable if it can be driven from any initial state to any desired final state, by proper variable inputs, in finite time \cite{lombardi2007controllability}. Most real systems are driven by nonlinear processes, but the controllability of nonlinear systems is in many aspects structurally similar to that of linear systems \cite{liu2011controllability}.
The linear and time-invariant (LTI) dynamics on a directed network can be described by:
\begin{equation}\label{equa1}
    \frac{d\mathbf{x}(t)}{dt}=\mathbf{A} \mathbf{x}(t)+\mathbf{B}\mathbf{u}(t),
\end{equation}
    where the $\textit{N} \times \text{1}$ vector $\mathbf{x}(t)=(x_1(t),x_2(t),...,x_N(t))^T$ denotes the state of the system with $N$ nodes at time \textit{t}. The weighted matrix $\mathbf{A}$ is an $\textit{N} \times \textit{N}$ matrix which describes the network topology and the interaction strength between the components. The $N \times M$ matrix $\mathbf{B}$ is the input matrix which identifies the $\textit{M} \le \textit{N}$ driver nodes controlled by outside input signals. The $\textit{M} \times \text{1}$ vector $\mathbf{u}(t)=(u_1(t),u_2(t),...,u_M(t))^T$ is the input signal vector. A driver node $j\in\{1,\dots,M\}$ has an input signal $u_j(t)$ which is externally fed into it.

The LTI system defined by Eq.~\eqref{equa1} is controllable if and only if the $\textit{N} \times \textit{NM}$ controllability matrix:
\begin{equation}
    \mathbf{C}=(\mathbf{B},\mathbf{AB},\mathbf{A}^{2}\mathbf{B},...,\mathbf{A}^{N-1}\mathbf{B}),
\end{equation}
    has full rank, i.e., $\text{rank}(\mathbf{C})=N$. This criterion is the so-called Kalman controllability rank condition~\cite{kalman1963mathematical}. The rank of the matrix $\mathbf{C}$ provides the dimension of the controllable subspace of the system. One therefore needs to find the right input matrix $\mathbf{B}$ consisting of a minimum number of driver nodes to ensure that the controllability matrix $\mathbf{C}$ has full rank.
    
\subsection{Driver nodes}
\begin{figure}
\includegraphics{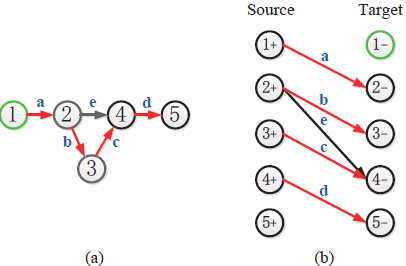}
\centering
\caption{Driver nodes and matching links (shown in red) in a directed network $G$. (a) An example network $G$ with $N=5$ nodes and $L=5$ directed links. Unmatched nodes are shown in green. (b) The corresponding bipartite graph with $2N$ nodes and $L$ links. By using the Hopcroft--Karp algorithm, a maximum set of matching links can be found in the bipartite graph. The target nodes of matching links are matched nodes. Other target nodes are unmatched nodes, which are also driver nodes.\label{fig:example_out_in}}
\end{figure}

Liu \emph{et al.} \cite{liu2011controllability} proved that the minimum number of driver nodes needed for structural controllability, where the input signals are injected to control the directed network, can be obtained through the ``maximum matching'' of the network. The source node of a directed link is defined as the node from which the link originates, while the target node is the node where the link terminates. A maximum matching of a directed network is a maximum set of links that do not share source or target nodes~\cite{yang2016mining}, which is illustrated in Fig.\ref{fig:example_out_in}(a). Such links are coined ``matching links''. Target nodes of matching links are matched nodes and the other nodes are unmatched nodes. For a given maximum matching, connecting driver nodes with unmatched nodes gives a minimum number of driver nodes $N_D$ needed for controlling the network.

A directed network $G$ with $N$ nodes and $L$ links can be converted into a bipartite graph $B_{N,N}$ with $2N$ nodes and $L$ links in order to find the maximum number of matching links, so as to determine the minimum number of driver nodes $N_D$ (see Fig.~\ref{fig:example_out_in}(b)). A maximum matching in a bipartite graph can be obtained efficiently by the Hopcroft--Karp algorithm~\cite{hopcroft1973n} when the original directed network is small. The unmatched nodes in a maximum matching constitute a minimum set of driver nodes. It is worth noting that a minimum set of driver nodes is not necessarily unique. The Hopcroft--Karp algorithm guarantees to return the minimum number of driver nodes to completely control the network. In addition, the computational complexity of the Hopcroft--Karp algorithm to find all driver nodes is $O(\sqrt{N}L)$.

As discussed above, the Hopcroft--Karp algorithm works efficiently when the network is small and sparse. However, in real life, this is seldom the case. When the network is large and dense, the Hopcroft--Karp algorithm may no longer be a viable efficient option in finding the number of driver nodes. Even with the rapid increase in computational power, the use of Hopcroft--Karp algorithm can be rendered ineffective if one tries to identify the sensitivity of the number of driver nodes with respect to several parameters characterizing the SSN topology. For instance, as we will see with Theorem 5.3, the number $n_D$ of driver nodes can have a very nonlinear, implicit and intricate relationship with the parameters defining the degree distribution. In such cases, performing a systematic sensitivity analysis of the dependence of $n_D$ with respect to these parameters using the Hopcroft--Karp algorithm would prove prohibitive. As an alternative, there exists a general expression for the minimum number $N_D$ of driver nodes obtained by using generating functions~\cite{newman2003structure}, which is also provided in~\cite{liu2011controllability}. However, this approach requires the knowledge of the closed-form degree distribution of the network. In the rest of this paper, we use this general expression to estimate the minimum number $N_D$ of driver nodes in the SSNs with regular out-degree distribution, and then deduce the general formula by considering the scenario when a fraction $p$ of the links are unavailable. We subsequently relax the condition that the out-degree is regular and look into networks with bi-modal out-degree distributions.

\section{Generating functions}
In a network, let $x$ denote the probability that a link is in state $X$. For example, $X$ can denote the weight of each link in a weighted network: $X$ can denote the existence of a link in an unweighted network. We assume that the states of links are independent from each other. Then, the probability that all the links of a node with degree $k$ are in state $X$ is $x^k$. Averaging this probability by the degree distribution of the network, we then obtain the probability that all the links of a randomly chosen node are in state $X$. According to the definition of the generating function \cite{newman2010networks}, this probability can be written as
\begin{equation}\label{G_x}
    G(x)=\sum^{\infty}_{k=0}p_kx^k,
\end{equation}
where $p_k$ is the probability that a randomly chosen node in the network has degree $k$.
Let $x=1$, then we obtain $G(1)=\sum^{\infty}_{k=0}p_k=1$. Besides, the average degree $\langle k\rangle$ of the network can be expressed as:
\begin{equation}\label{average_k}
    \langle k\rangle=G^{'}(1)=\sum^{\infty}_{k=0}kp_k.
\end{equation}

%\RB{STOPPED HERE}
Considering the degree of the node reached by following a randomly chosen link is $k$, the probability that all the other links of this node are in state $X$ is $x^{k-1}$. The distribution of the degrees of the nodes reached by following a randomly chosen link is called the excess degree distribution $q_k$, which depends on the degree distribution $p_k$. Note that the larger $p_k$ is, the larger $q_k$ is. Furthermore, following a link, it is easier to reach a node with larger $k$. Hence, we have
\begin{equation}\label{q_k}
    q_k\propto kp_k.
\end{equation}
The normalized distribution $q_k$ is
\begin{equation}\label{q_k_2}
    q_k=\frac{kp_k}{\sum^{\infty}_{k=0}kp_k}=\frac{kp_k}{\langle k\rangle}.
\end{equation}
Thus, the probability that all the other links of a node reached by following a randomly chosen link are in state $X$ is given by
\begin{equation}\label{H_x}
    H(x)=\sum^{\infty}_{k=1}q_kx^{k-1}=\sum^{\infty}_{k=1}\frac{kp_k}{\langle k\rangle}x^{k-1}=\frac{G^{'}(x)}{G^{'}(1)}.
\end{equation}
It must be highlighted that all these functions are based on the assumption that the states of links are independent from each other \cite{newman2003structure}.

\section{SSNs with $k$-regular out-degree}
\subsection{Fraction of driver nodes in SSNs with $k$-regular out-degree}

It is shown in Liu et al \cite{liu2011controllability} that the minimum number of driver nodes can be obtained by using the following set of generating functions 
\begin{align}
    G_{\text{out}}(x)&=\sum^{\infty}_{k_{\text{out}}=0}P_{\text{out}}(k_{\text{out}})x^{k_{\text{out}}},\label{Gout}\\
    G_{\text{in}}(x)&=\sum^{\infty}_{k_{\text{in}}=0}P_{\text{in}}(k_{\text{in}})x^{k_{\text{in}}},\label{Gin}\\
    H_{\text{out}}(x)&=\sum^{\infty}_{k_{\text{out}}=1}\frac{k_{\text{out}}P_{\text{out}}({k_{\text{out}}})}{\langle k_{\text{out}}\rangle}x^{k_{\text{out}}-1},\label{Hout}\\
    H_{\text{in}}(x)&=\sum^{\infty}_{k_{\text{in}}=1}\frac{k_{\text{in}}P_{\text{in}}({k_{\text{in}}})}{\langle k_{\text{in}}\rangle}x^{k_{\text{in}}-1},\label{Hin}
\end{align}
where $P_{\text{out}}(\cdot)$ and $P_{\text{in}}(\cdot)$ denote the probability distribution function of the out-degree and in-degree, respectively, and $\langle k_{\text{out}}\rangle$ and $\langle k_{\text{in}}\rangle$ denote the average out-degree and in-degree, respectively.

Using those generating functions, the general expression for the minimum fraction $N_D$ of driver nodes derived by Liu et al.~\cite{liu2011controllability} reads
\begin{equation}\label{nD}
\begin{split}
    n_D=\frac{N_D}{N}=
    \frac{1}{2}\{G_{\text{in}}(w_2)+G_{\text{in}}(1-w_1)-2+G_{\text{out}}(\hat{w}_2)+G_{\text{out}}(1-\hat{w}_1)+\\
    k(\hat{w}_1(1-w_2)+w_1(1-\hat{w}_2))\},
\end{split}
\end{equation}
where $w_1, w_2, \hat{w}_1$ and $\hat{w}_2$ satisfy
\begin{align}
w_1&=H_{\text{out}}(\hat{w}_2),    \label{eq1}\\
w_2&=1-H_{\text{out}}(1-\hat{w}_1),    \label{eq2}\\
\hat{w}_1&=H_{\text{in}}(w_2),    \label{eq3}\\
\hat{w}_2&=1-H_{\text{in}}(1-w_1).    \label{eq4}
\end{align}
By construction, the out-degree distribution for the SSN suggested in~\cite{Komareji} is a Dirac delta function, i.e. 
\begin{equation}\label{Pout}
    P_{\text{out}}(k_{\text{out}})=\delta(k-k_{\text{out}}),
\end{equation}
where $k$ is the fixed out-degree for every node. Thus, the average out-degree $\langle k_{out} \rangle$ equals the out-degree $k$ of each node. 
It is also shown in \cite{Komareji} that, for sufficiently large SSNs, the in-degree distribution closely resembles a Poisson distribution, with average $k$, i.e.
\begin{equation}\label{Pin}
    P_{\text{in}}(k_{\text{in}})=\frac{k^{k_{\text{in}}}}{k_{\text{in}}!}e^{-k}.
\end{equation}
Using the degree distributions in Eqs.~\eqref{Gout}--\eqref{Hin} it follows
\begin{align}
    G_{\text{out}}(x)&=x^{k},\label{Gout2}\\
    G_{\text{in}}(x)&=e^{-k(1-x)},\label{Gin2}\\
    H_{\text{out}}(x)&=x^{k-1},\label{Hout2}\\
    H_{\text{in}}(x)&=e^{-k(1-x)}.\label{Hin2}
\end{align}
Therefore, the parameters $w_1, w_2, \hat{w}_1$ and $\hat{w}_2$ satisfy
\begin{align}
w_1&=\hat{w}_2^{k-1},\label{eq1B}\\    
w_2&=1-(1-\hat{w}_1)^{k-1},    \label{eq2B}\\
\hat{w}_1&=e^{-k(1-w_2)},    \label{eq3B}\\
\hat{w}_2&=1-e^{-kw_1}.    \label{eq4B}
\end{align}
For the trivial case $k=0$, it is easy to see that the above set of equations leads to $n_D=1$, i.e. all agents in the swarm need to be controlled, which makes sense because the out-degree of every node is $0$ in this case. Also, for the case $k=1$, Eqs.~\eqref{eq1B}--\eqref{eq4B} are solved for $w_1 = 1, w_2 = 0, \hat{w}_1 = e^{-1}$ and $\hat{w}_2=1-e^{-1}$. Hence, for $k=1$, it holds that $n_D=e^{-1}$.

\begin{figure}[htbp!]\centering
\centering
\includegraphics{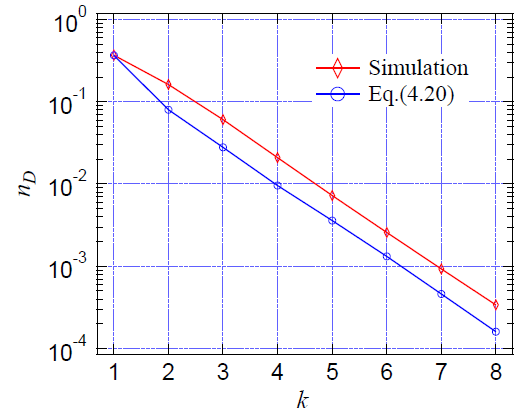}
\caption{Fraction of driver nodes $n_D$ for different values of the out-degree $k$: Eq.~\eqref{nD_wrong} versus simulation results.}\label{fig:wrong}
\end{figure}

For the case $k>1$, Komareji \& Bouffanais~\cite{Komareji} argue that the smallest solution of the pair of Eqs.~\eqref{eq1B} and \eqref{eq4B} is given by
$w_1 = \hat{w}_2 = 0$, and assuming that $w_1$ and $\hat{w}_2$ are indeed zero, the following expression for the fraction of driver nodes is derived:
\begin{equation}\label{nD_wrong}
    n_D=\frac{1}{2}\{(1-e^{-k(1-w_2)})^k - 1 + e^{-k(1-w_2)} + k(1-w_2)e^{-k(1-w_2)}\},
\end{equation}
where $w_2$ is the solution of the implicit equation
\begin{equation}\label{w2_eq}
    1-w_2 = (1-e^{-k(1-w_2)})^{k-1}.
\end{equation}
From Eq.~\eqref{nD_wrong} the asymptotic behaviour of $n_D$ for large $k$ can also be derived:
\begin{equation}\label{approx_wrong}
    n_D \approx \frac{1}{2}e^{-k}.
\end{equation}
However, upon simulation of SSNs, determining the fraction of driver nodes by applying the maximum matching algorithm, as described in~\cite{liu2011controllability}, we found a discrepancy between Eq.~\eqref{nD_wrong} and the simulation results shown in Fig.~\ref{fig:wrong}.
We generate $10000$ directed networks with $N=10000$ nodes each having an out-degree $k$ whose value ranges from 1 to 8. The fraction $n_D$ of driver nodes is the average fraction of driver nodes over $10000$ networks for each out-degree $k$. As shown in Fig.~\ref{fig:wrong}, the result from Eq.~\eqref{nD_wrong} fits well with the simulation result at $k=1$. However, the difference between Eq.~\eqref{nD_wrong} and simulation results are obvious for other values of $k$. For example, at all 
points $k>1$, the results from the simulation are about two times the results given by Eq.~\eqref{nD_wrong}.   

The discrepancy is due to the assumption that one can choose the solution of Eq.~\eqref{eq1B} and Eq.~\eqref{eq4B} given by $w_1 = \hat{w}_2 = 0$. One can also argue that the pair Eq.~\eqref{eq1B} and Eq.~\eqref{eq4B} is equivalent to the pair Eq.~\eqref{eq2B} and Eq.~\eqref{eq3B}. If we assume
\begin{align}
w_1&=1-w_2,\\
\hat{w}_2&=1-\hat{w}_1,
\end{align}
then the pair of equations Eq.~\eqref{eq2B}--Eq.~\eqref{eq3B} follows from the pair of equations Eq.~\eqref{eq1B}--Eq.~\eqref{eq4B}.
As a result, applying Eq.~\eqref{nD} leads to the following expression for the fraction of driver nodes:
\begin{equation}\label{nD_correct}
    n_D=((1-e^{-k(1-w_2)})^k - 1 + e^{-k(1-w_2)} + k(1-w_2)e^{-k(1-w_2)}),
\end{equation}
where $w_2$ is still the solution of Eq.~\eqref{w2_eq}.

The asymptotic behavior of $n_D$ for large $k$ becomes:
\begin{equation}\label{nD_approx_correct}
    n_D \approx e^{-k}.
\end{equation}
Note that Eq.~\eqref{nD_correct} also holds for $k=1$, another indication of its correctness. 

Table 1 shows the comparison between the approximations in Eqs.~\eqref{nD_correct} and \eqref{nD_approx_correct} and the simulations.

%\begin{figure}[htbp]
%\centering
%\includegraphics[width=0.8\textwidth]{figures/spoke_clustering_sweep.eps}                   
%\includegraphics[width=0.4\textwidth]{final_figures/Table1.png}
%\caption{Comparing Eqs.\eqref{nD_correct}-\eqref{nD_approx_correct} with simulation results.}\label{Table1}
%\end{figure}

\begin{table}[ht]
\centering 
\caption{Comparison of Eqs.~\eqref{nD_correct}--\eqref{nD_approx_correct} with simulation results.}
\begin{tabular}{l|l|l|l|l|l}
\hline
\multirow{2}{*}{\textbf{$k$}} &
\multicolumn{2}{|c|}{\textbf{Eq.~\eqref{nD_correct}}} &
\multicolumn{2}{|c|}{\textbf{Eq.~\eqref{nD_approx_correct}}} &
\multirow{2}{*}{\textbf{Simulations}} 
\\
\cline{2-3}  %  \cline用于画横线 \cline{i-j}表示从第i列画到第j列
\cline{4-5}
& \textbf{value} & $r$ & \textbf{value} & $r$ \\

\hline
1     & 0.367879       &0.0079\%       & 0.367879    &0.0079\%     &0.36782\\ \hline
2     & 0.161903      &0.40\%         & 0.135335       &16.07\%        &0.162003\\ \hline
3    & 0.060759        &0.29\%       & 0.049787         &17.82\%      &0.06068\\ \hline
4    & 0.020916        &0.28\%       & 0.018316       &12.18\%        &0.020943\\ \hline
5   & 0.007262        &0.93\%       & 0.006738       &6.35\%        &0.007221\\ \hline
6   & 0.002578    &0.23\%          & 0.002479     &4.06\%          &0.002561\\ \hline
7 & 0.00093       &2.76\%        & 0.000912       &0.77\%       &0.000929\\ \hline
8   & 0.000339      &5.93\%         & 0.000335      &4.69\%        &0.000346\\ \hline

\end{tabular}
\end{table}

Like previously, we generate $10000$ directed networks with $N=10000$ for each out-degree $k$ whose value ranges from 1 to 8. The fraction of driver nodes $n_D$ is the average fraction of driver nodes in $10000$ networks. Then we calculate the analytical results from Eq.~\eqref{nD_correct} and Eq.~\eqref{nD_approx_correct} and also the corresponding absolute relative error $r$. As shown in Table 1, the absolute relative errors of our approximation are less than 1\% for $k$ from 1 to 6. For the case where $k=7$ and $k=8$, the absolute relative errors are still small---less than 6\%. When the values of $k$ are small, the absolute relative errors of Eq.~\eqref{nD_approx_correct} are large.

We conclude from Table 1 that the simulations are an excellent fit for our approximation in Eq.~\eqref{nD_correct}. Also, the asymptotic approximation Eq.~\eqref{nD_approx_correct} is increasingly accurate for increasing $k$. 

\subsection{Fraction of driver nodes under random link failures}
In this section, we generalize the results of the previous section by considering SSNs with $k$-regular out-degree, but now we assume that a fraction $p$ of the links is removed at random. This assumption is in accordance with some real-life scenarios, such as the communication disconnection between robots in swarm robotic networks because of the limited range of communication.

In what follows, we show that the analysis that led to our implicit approximations is still valid and applicable for this case.
A crucial step is to find expressions for the generating functions Eqs.~\eqref{Gout}--\eqref{Hin} for this specific case involving a fraction of link failures.

The following lemma is instrumental in establishing the key results for this case---see \cite{vanmieghem_2014} which gives an expression for the degree distribution, after removing $m$ links uniformly at random.

\begin{lemma}\label{L1}
After removing $m$ links in a uniform and random way from a network $G_0(N,L)$, with degree distribution $Pr[D_{G_0}=j]$, the degree distribution $Pr[D_{G}=i]$ of the new network $G$ satisfies:
\begin{equation}
Pr[D_G=i]=(1-p)^i\sum^{N-1}_{j=i}\binom{j}{i}p^{j-i}Pr[D_{G_0}=j],
\end{equation}
\end{lemma}
where $p=\frac{m}{L}$ denotes the fraction of removed links in the original network $G_0$.

\begin{theorem}\label{T2}
Consider a directed network with a $k$-regular out-degree and a Poisson in-degree distribution with average $k$. Upon removing uniformly and at random a fraction $p$ of the links, the generating functions $\bar{G}_{\text{\em out}}(x)$ and $\bar{G}_{\text{\em in}}(x)$ of the out- and in-degree, respectively, satisfy
\begin{equation}\label{Gout_bar}
    \bar{G}_{\text{\em out}}(x)=(p + (1-p)x)^k,
\end{equation}
\begin{equation}\label{Gin_bar}
    \bar{G}_{\text{\em in}}(x)=e^{-k(1-p)(1-x)}.
\end{equation}
\end{theorem}

The proof of Theorem~\ref{T2} is given in Appendix A.
Note that for the case without link removals, i.e. $p=0$, Eqs.~\eqref{Gout_bar}--\eqref{Gin_bar} reduce to Eqs.~\eqref{Gout2}--\eqref{Gin2}. Also, we can deduce from Eqs.~\eqref{Gout_bar}--\eqref{Gin_bar} directly that both the average out- and in-degree after link removals, which is denoted by $\bar{k}$, equal
\begin{equation}
    \bar{k}=k(1-p).
\end{equation}

\begin{theorem}\label{T3}
Consider a directed network with a $k$-regular out-degree and a Poisson in-degree with average $k$. Then, after removing uniformly at random a fraction $p$ of the links, the generating functions $\bar{H}_{\text{\em out}}(x)$ and $\bar{H}_{\text{\em in}}(x)$ of the excess out- and in-degree, respectively, satisfy
\begin{equation}\label{Hout_bar}
    \bar{H}_{\text{\em out}}(x)=(p + (1-p)x)^{k-1}
\end{equation}
\begin{equation}\label{Hin_bar}
    \bar{H}_{\text{\em in}}(x)=e^{-k(1-p)(1-x)}
\end{equation}
\end{theorem}

The proof of Theorem \ref{T3} is given in Appendix A.
Note that for the case without link removals, i.e. $p=0$, Eqs.~\eqref{Hout_bar}--\eqref{Hin_bar} reduce to Eqs.~\eqref{Hout2}--\eqref{Hin2}.
\newline
\newline
The results in Theorems \ref{T2} and \ref{T3} can also be directly deduced by using a result from~\cite{vanmieghem_2014}: if the generating function for the degree distribution for a network is given by $G(x)$, then the generating function $\bar{G}(x)$ for the resulting network after a fraction $p$ of links are randomly removed, satisfies $\bar{G}(x)=G(p+(1-p)x)$. In addition, Theorem \ref{T3} can also be established directly by applying Eq.~\eqref{H_x} to Eqs.~\eqref{Gout_bar}--\eqref{Gin_bar}.

We are now in a position to state the following result.

\begin{theorem}\label{T4}
Consider a directed network with a $k$-regular out-degree and a Poisson in-degree distribution with average $k$. Then, after removing uniformly at random a fraction $p$ of its links, the fraction of the minimum number of driver nodes is given by
\begin{equation}\label{nD_bar}
\begin{split}
    n_D=(p+(1-p)(1-e^{-k(1-p)(1-w_2)}))^k - 1 + e^{-k(1-p)(1-w_2)} +\\ k(1-p)(1-w_2)e^{-k(1-p)(1-w_2)},
    \end{split}
\end{equation}
where $w_2$ satisfies
\begin{equation}\label{w2_eq_bar}
    1-w_2 = (p+(1-p)(1-e^{-k(1-p)(1-w_2)}))^{k-1}.
\end{equation}
The asymptotic behavior of $n_D$ for large $k$ is given by
\begin{equation}\label{nD_bar_approx}
    n_D \approx e^{-k(1-p)}.
\end{equation}
\end{theorem}
It is worth noting that for the particular case without link removals, i.e. $p=0$, Eqs.~\eqref{nD_bar}--\eqref{nD_bar_approx} reduce to Eqs.~\eqref{nD_correct}-\eqref{w2_eq}-\eqref{nD_approx_correct}, respectively. 
The proof of Theorem \ref{T4} is given in Appendix A.

Table 2 shows the comparison between the approximations in Eqs.~\eqref{nD_bar} and \eqref{nD_bar_approx} and simulations, for the cases $p=0.2$ and $p=0.5$. Specifically, we generated $1000$ directed networks with $N=10000$ with out-degree $k$, where $k \in \{1,2,3,4,5,6,7,8\}$. For each network with the same out-degree $k$, we randomly removed a fraction $p$ of links and get the value of $n_D$, and then repeat this process one thousand times. Thus, the fraction of driver nodes $n_D$ for a combination $(k,p)$ is the average fraction of driver nodes in $10^{6}$ realizations.

As shown in Table 2, the absolute relative errors $r$ of our approximation Eq.~\eqref{nD_bar} are small---less than 4\% for any $k$ value when $p=0.2$ or $p=0.5$. In contrast, the relative errors of the asymptotic approximation Eq.~\eqref{nD_bar_approx} are large for most cases.

%\begin{figure}[htbp!]
%\centering
%\includegraphics[width=0.8\textwidth]{figures/spoke_clustering_sweep.eps}                   
%\includegraphics[width=0.8\textwidth]{final_figures/Table2.png}
%\caption{Comparing Eqs.\eqref{nD_bar}-\eqref{nD_bar_approx} with simulation results.}\label{Table2}
%\end{figure}

\begin{table}[ht]
\centering
\caption{Comparison of Eqs.~\eqref{nD_bar}--\eqref{nD_bar_approx} with simulation results.}
\label{Table2}
\resizebox{\textwidth}{!}{

\begin{tabular}{l|l|l|l|l|l|l|l|l|l|l}
\hline
\multirow{2}{*}{$k$} &
\multicolumn{4}{|c|}{\textbf{Eq.~\eqref{nD_bar}}} &
\multicolumn{4}{|c|}{\textbf{Eq.~\eqref{nD_bar_approx}}} &
\multicolumn{2}{|c}{\textbf{Simulations}}
\\
\cline{2-3}  %  \cline用于画横线 \cline{i-j}表示从第i列画到第j列
\cline{4-5}
\cline{6-7}  %  \cline用于画横线 \cline{i-j}表示从第i列画到第j列
\cline{8-9}
\cline{10-11}
& $p=0.2$ & $r$ & $p=0.5$ & $r$ & $p=0.2$ & $r$ & $p=0.5$ & $r$ & $p=0.2$ &  $p=0.5$\\

\hline
1     & 0.442926       &1.41\%       & 0.584101    &3.72\%      &0.449329   &0.019\%        & 0.606531   &0.021\%  & 0.449321    &0.606622\\ \hline
2     & 0.238827      &0.30\%          & 0.410116       &0.12\%         &0.201897  &15.21\%    & 0.367879      &10.20\%  & 0.238905    &0.410229\\ \hline
3    & 0.116278        &0.30\%        & 0.279218        &0.24\%       &0.090718  &21.75\%        & 0.22313       &19.89\%  & 0.116176    &0.279108\\ \hline
4    & 0.050341        &0.38\%        & 0.183439       &0.19\%         &0.040762  &18.72\%        & 0.135335    &26.08\%  & 0.050167    &0.183421\\ \hline
5   & 0.021143        &0.96\%        & 0.112696       &0.29\%         &0.018316  &12.53\%        & 0.082085       &26.95\%  & 0.021215    &0.112680\\ \hline
6   & 0.009002   &0.13\%           & 0.065394    &0.55\%           &0.00823   &8.70\%           & 0.049787     &23.45\%  & 0.009041   &0.065339\\ \hline
7 & 0.003902       &0.20\%         & 0.037384       &1.18\%         &0.003698   &5.41\%         & 0.030197       &18.92\%  & 0.003915    &0.03736\\ \hline
8   & 0.001714      &2.50\%          & 0.021502      &0.20\%         &0.001662  &5.5\%          & 0.018316      &14.65\%  & 0.001706    &0.021533\\ \hline

\end{tabular}
}
\end{table}

We conclude from Table 2 that the simulations yield an excellent fit with our approximation Eq.~\eqref{nD_bar}. Unsurprisingly, the asymptotic approximation Eq.~\eqref{nD_bar_approx} is increasingly accurate for increasing $k$.

\begin{figure}[htbp!]
\centering
\includegraphics{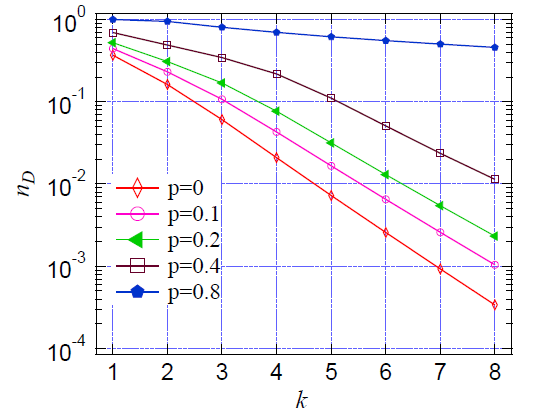}
\caption{Fraction of driver nodes as function of the out-degree $k$ for several values of the fraction of removed links $p$.}\label{fig:3}
\end{figure}

Finally, Fig. \ref{fig:3} shows the fraction of driver nodes $n_D$ as a function of the out-degree $k$ for several values of $p$. The value of $n_D$ decreases as the degree of networks increases for a specific $p$. Note that for the same $k$ value, a larger value of $p$ leads to a larger value of $n_D$.

\section{SSNs with a bi-modal out-degree}
\subsection{Fraction of driver nodes in SSNs with a bi-modal out-degree}
In this section, we generalize the results of one of the previous sections by considering SSNs with a bi-modal out-degree distribution, i.e. we assume that for a fraction $\alpha$ of nodes the out-degree is $k_1$, while for the remaining fraction $1-\alpha$ of nodes, the out-degree equals $k_2$. We will assume $k_1 \neq k_2$ and both $k_1$ and $k_2$ are strictly larger than $0$.

\begin{theorem}\label{T5}
Consider a directed network with a bi-modal out-degree distribution $\alpha\delta(k_{\text{\emph{out}}}-k_1)+\alpha\delta(k_{\text{\emph{out}}}-k_2)$, with average out-degree 
\begin{equation}\label{k_bi}
   k  =\alpha k_1 + (1-\alpha) k_2,
\end{equation}
and a Poisson in-degree distribution with average $k$.
The generating functions $\hat{G}_{\text{\emph{out}}}(x)$ and $\hat{G}_{\text{\emph{in}}}(x)$ of the out- and in-degree, respectively, satisfy
\begin{equation}\label{Gout_bi}
    \hat{G}_{\text{\emph{out}}}(x)=\alpha x^{k_1}+(1-\alpha)x^{k_2},
\end{equation}
\begin{equation}\label{Gin_bi}
    \hat{G}_{\text{\emph{in}}}(x)=e^{-k(1-x)}.
\end{equation}
\end{theorem}

The proof of Theorem \ref{T5} is given in Appendix B.

\begin{theorem}\label{T6}
Consider a directed network with bi-modal out-degree $\alpha\delta(k_{\text{\emph{out}}}-k_1)+\alpha\delta(k_{\text{\emph{out}}}-k_2)$, with average out-degree 
\begin{equation}
    k =\alpha k_1 + (1-\alpha) k_2
\end{equation}
and a Poisson in-degree distribution with average $k$. Then, the generating functions $\hat{H}_{\text{\emph{out}}}(x)$ and $\hat{H}_{\text{\emph{in}}}(x)$ of the excess out- and in-degree, respectively, satisfy
\begin{align}
    \hat{H}_{\text{\emph{out}}}(x)&=\frac{\alpha k_1 x^{k_1-1}+(1-\alpha) k_2 x^{k_2-1}}{k},\label{Hout_bi}\\
    \hat{H}_{\text{\emph{in}}}(x)&=e^{-k(1-x)}.\label{Hin_bi}
\end{align}
\end{theorem}

The proof of Theorem \ref{T6} is given in Appendix B. The proof also can be established by applying Eq.~\eqref{H_x} directly to Eqs.~\eqref{Gout_bi}--\eqref{Gin_bi}. Note for the case $k_1=k_2=k$, where the out-degree reduces to a Dirac function, Eqs.~\eqref{Gout_bi}--\eqref{Hin_bi} reduce to Eqs.~\eqref{Gout2}--\eqref{Hin2}.

\begin{theorem}\label{T7}
Consider a directed network with a bi-modal out-degree $\alpha\delta(k_{\text{\emph{out}}}-k_1)+\alpha\delta(k_{\text{\emph{out}}}-k_2)$, with average out-degree $k =\alpha k_1 + (1-\alpha) k_2$ and a Poisson in-degree with average $k$. Then, the fraction of minimum number of driver nodes is given by
\begin{equation}\label{nD_bi}
    n_D=\alpha(1-e^{-k(1-w_2)})^{k_1}+
(1-\alpha)(1-e^{-k(1-w_2)})^{k_2}-1+e^{-k(1-w_2)}+ke^{-k(1-w_2)}(1-w_2), \end{equation}
where $w_2$ satisfies
\begin{equation}\label{w2_eq_bi}
    1-w_2 = \frac{\alpha k_1 (1-e^{-k(1-w_2)})^{k_1-1}+(1-\alpha) k_2 (1-e^{-k(1-w_2)})^{k_2-1}}{k}.
\end{equation}
The asymptotic behaviour of $n_D$ for large $k$ is given by
\begin{equation}\label{nD_approx_bi}
    n_D \approx e^{-k}.
\end{equation}
\end{theorem}
Note for the case $k_1=k_2=k$, where the out-degree reduces to a Dirac function, Eqs.~\eqref{nD_bi}--\eqref{nD_approx_bi} reduce to Eqs.~\eqref{nD_correct}--\eqref{w2_eq}--\eqref{nD_approx_correct}, respectively. It is worth noting that Eqs.~\eqref{nD_bi}--\eqref{w2_eq_bi} reveal a complex dependency of $n_D$ with respect to the parameters characterizing the degree distribution, namely $(\alpha,k_1,k_2)$. This is particularly apparent given the implicit nature of Eq.~\eqref{w2_eq_bi}. In such cases, identifying the intricate dependency of $n_D$ with those parameters using the Hopcroft--Karp algorithm would require an impractical brute-force approach.
\newline

The proof of Theorem \ref{T7} is given in Appendix B.
\newline

Table 3 shows the comparison between the approximations in Eqs.~\eqref{nD_bi} and \eqref{nD_approx_bi} and simulations.

We generate $10000$ directed networks with $N=10000$ for each out-degree combination $(k_1,k_2,\alpha)$ and obtain the average fraction $n_D$ of driver nodes. As shown in Table 3, the absolute relative errors $r$ of our approximation Eq.~\eqref{nD_bi} are small, thereby indicating a good fit with simulations.
The absolute relative errors of Eq.~\eqref{nD_approx_bi} are larger, especially for small average degree $k = \alpha k_1 + (1-\alpha) k_2$.
%\begin{figure}[htbp!]
%\centering
%\includegraphics[width=0.8\textwidth]{figures/spoke_clustering_sweep.eps}                   
%\includegraphics[width=0.8\textwidth]{final_figures/Table3.png}
%\caption{Comparing Eqs.\eqref{nD_bi}-\eqref{nD_approx_bi} with simulation results.}\label{Table3}
%\end{figure}

\begin{table}[ht]
\centering
\caption{Comparing Eqs.~\eqref{nD_bi}--\eqref{nD_approx_bi} with simulation results.}
\label{tab:outiner_results_simappann}
\resizebox{\textwidth}{!}{
\begin{tabular}{l|l|l|l|l|l|l|l|l}
\hline
\multirow{2}{*}{$k_1$} &
\multirow{2}{*}{$k_2$} &
\multirow{2}{*}{$k$} &
\multirow{2}{*}{$\alpha$} &
\multicolumn{2}{|c|}{\textbf{Eq.~\eqref{nD_bi}}} &
\multicolumn{2}{|c|}{\textbf{Eq.~\eqref{nD_approx_bi}}} &
\multirow{2}{*}{\textbf{Simulation}} 
\\
\cline{5-8}  %  \cline用于画横线 \cline{i-j}表示从第i列画到第j列
\cline{6-7}
& & & &\textbf{value} & $r$ & \textbf{value} & $r$ \\

\hline
1  &3 &2.5 &0.25 & 0.107746       &0.51\%       & 0.082085    &23.43\%     &0.107795\\ \hline
1  &3 &2 &0.5  & 0.183062      &0.020\%         & 0.135335       &26.09\%        &0.181395\\ \hline
1   &3 &1.5 &0.75 & 0.273670        &0.040\%       & 0.223130         &18.44\%      &0.273455\\ \hline
2  &4 &3.5 &0.25  & 0.036402        &0.56\%       & 0.030197       &16.58\%        &0.036705\\ \hline
2  &4 &3 &0.5 & 0.063648       &0.27\%       & 0.049787      &21.57\%        &0.06352\\ \hline
2  &4 &2.5 &0.75 & 0.106955    &0.25\%          & 0.082085     &23.44\%          &0.106735\\ \hline
2 &6 &5 &0.25 & 0.007355       &1.04\%        & 0.006738       &9.30\%        &0.007315\\ \hline
2  &6 &4 &0.5 & 0.022172     &0.76\%         & 0.018316      &16.76\%        &0.022335\\ \hline
2  &6 &3 &0.75 & 0.071349      &0.19\%       & 0.049787    &30.09\%     &0.071875\\ \hline
2  &8 &6.5 &0.25  & 0.001555      &3.81\%         & 0.001503       &0.33\%        &0.001595\\ \hline
2  &8 &5  &0.5 & 0.007556       &2.20\%       & 0.006738         &8.86\%      &0.007745\\ \hline
2  &8 &3.5 &0.75  & 0.045382        &0.35\%       & 0.030197       &33.69\%        &0.04665\\ \hline
4  &6 &5.5 &0.25 & 0.004324        &0.68\%       & 0.004087       &4.84\%        &0.004362\\ \hline
4  &6 &5 &0.5 & 0.007293    &0.97\%          & 0.006738     &6.71\%          &0.007181\\ \hline
4 &6 &4.5 &0.75 & 0.012357      &1.40\%        & 0.011109       &8.34\%        &0.01228\\ \hline
4  &8 &7 &0.25 & 0.000931     &3.22\%         & 0.000912      &5.20\%        &0.000962\\ \hline
4  &8 &6 &0.5 & 0.002593      &4.18\%         & 0.002479      &4.36\%        &0.002706\\ \hline
4  &8 &5 &0.75 & 0.007354      &1.17\%         & 0.006738      &7.56\%        &0.007269\\ \hline
\end{tabular}
}
\end{table}

We conclude from Table 3 that the simulations constitute a very good fit with our approximation Eq.~\eqref{nD_bi}. Also, the asymptotic approximation Eq.~\eqref{nD_approx_bi} is increasingly accurate for increasing $k$.

\subsection{Fraction of driver nodes under random link failures}
In this section, we generalize the results of the previous section by considering again
SSNs with a bi-modal out-degree, but now we assume that a fraction $p$ of the links is removed at random. We show that the analysis that led to our implicit approximations can also be conducted for this case. Similar to the case with a regular out-degree, a crucial step is to find expressions for the generating functions Eqs.~\eqref{Gout}--\eqref{Hin} for this particular case.

Based on Lemma 1, we get:
\newline

\begin{theorem}\label{T8}
Consider a directed network with a bi-modal out-degree $\alpha\delta(k_{\text{\emph{out}}}- k_1) + (1-\alpha)\delta(k_{\text{\emph{out}}}- k_2)$, with average out-degree \\
\begin{equation}\label{eqkbi}
    k =\alpha{k_1}+(1-\alpha)k_2
\end{equation}
and a Poisson in-degree with average $k$. Then, after removing uniformly at random a fraction $p$ of the links, the generating functions $\tilde{G}_{\text{\emph{out}}}(x)$ and $\tilde{G}_{\text{\emph{in}}}(x)$ of the out- and in-degree, respectively, satisfy
\begin{align}
    \tilde{G}_{\text{\emph{out}}}(x)&=\alpha{(p+(1-p)x)^{k_1}}+(1-\alpha){(p+(1-p)x)^{k_2}}, \label{eqGbarout_remove}\\
    \tilde{G}_{\text{\emph{in}}}(x)&=e^{-k(1-p)(1-x)}.\label{eqGbarin_remove}
\end{align}
\end{theorem}
By applying the generating function $\bar{G}(x)$ for the resulting network after a fraction $p$ of links are randomly removed \cite{vanmieghem_2014}, the theorem also follows directly from $\tilde{G}_{\text{out}}(x)=\hat{G}_{\text{out}}(p+(1-p)x)$ and $\tilde{G}_{\text{in}}(x)=\hat{G}_{\text{in}}(p+(1-p)x)$. Note that for the case without link removals, i.e. $p = 0$, Eqs.~\eqref{eqGbarout_remove}--\eqref{eqGbarin_remove} reduce to Eqs.~\eqref{Gout_bi}--\eqref{Gin_bi}. Also, we can deduce from Eqs.~\eqref{eqGbarout_remove}--\eqref{eqGbarin_remove} directly that both the average out- and in-degree after link removals, which we denote by $\tilde{k}$, and satisfies
\begin{equation}\label{eqktilde}
    \tilde{k}=k(1-p).
\end{equation}

\begin{theorem}\label{T9}
Consider a directed network with a bi-modal out-degree $\alpha\delta(k_{\text{out}}- k_1) + (1-\alpha)\delta(k_{\text{out}}- k_2)$, with average out-degree
\begin{equation}\label{eqk_bi}
    k =\alpha{k_1}+(1-\alpha)k_2,
\end{equation}
and a Poisson in-degree with average $k$. Then, after removing uniformly at random a fraction $p$ of the links, the generating functions $\bar{H}_{\text{\emph{out}}}(x)$ and $\bar{H}_{\text{\emph{in}}}(x)$ of the excess out- and in-degree, respectively, satisfy
\begin{align}
    \tilde{H}_{\text{\emph{out}}}(x)&=\frac{\alpha{k_1(p+(1-p)x)^{k_1-1}}+(1-\alpha){k_2(p+(1-p)x)^{k_2-1}}}{k}, \label{eqHbarout_removed}\\
    \tilde{H}_{\text{\emph{in}}}(x)&=e^{-k(1-p)(1-x)}. \label{eqHbarin_removed}
\end{align}
\end{theorem}
The proof of Theorem \ref{T9} can readily be obtained by combining the proofs of Theorems \ref{T3} and \ref{T6}. By applying the generating function $\bar{G}(x)$ for the resulting network after a fraction $p$ of links are randomly removed \cite{vanmieghem_2014}, the theorem also follows directly from $\tilde{H}_{\text{out}}(x)=\hat{H}_{\text{out}}(p+(1-p)x)$ and $\tilde{H}_{\text{in}}(x)=\hat{H}_{\text{in}}(p+(1-p)x)$. Note that for the case without link removals, i.e. $p = 0$, Eqs.~\eqref{eqHbarout_removed}--\eqref{eqHbarin_removed} reduce to Eqs.~\eqref{Hout_bi}--\eqref{Hin_bi}.
\newline

After obtaining expressions for all required generation functions, we are now in the position to state the following result.
\newline

\begin{theorem}\label{T10}Consider a directed network with a bi-modal out-degree $\alpha\delta(k_{\text{out}}- k_1) + (1-\alpha)\delta(k_{\text{out}}- k_2)$, with average out-degree
$k =\alpha{k_1}+(1-\alpha)k_2$ and a Poisson in-degree with average $k$. Then, after removing uniformly at random a fraction $p$ of the links,  the fraction of minimum number of driver nodes is given by:
\begin{equation}\label{eq51}
\begin{split}
    n_D=\alpha{(p+(1-p)(1-e^{-k(1-\omega_2)}))}^{k_1}+(1-\alpha)(p+(1-p)(1-e^{-k(1-\omega_2)}))^{k_2}\\-1+e^{-k(1-p)(1-\omega_2)}+k(1-p)e^{-k(1-\omega_2)}(1-\omega_2)
\end{split}    
\end{equation}
where $\omega_2$ satisfies
\begin{equation}\label{eq52}
\begin{split}
    1-\omega_2=\\
    \frac{\alpha{k_1}(p+(1-p)(1-e^{-k(1-p)(1-\omega_2)}))^{k_1-1}+(1-\alpha)k_2(p+(1-p)(1-e^{-k(1-p)(1-\omega_2)}))^{k_2-1}}{k}.
\end{split}
\end{equation}
The asymptotic behaviour of $n_D$ for large k is given by
\begin{equation}\label{eq53}
    n_D\approx{e^{-k(1-p)}}.
\end{equation}
\end{theorem}
For the case without link removals, i.e. $p = 0$, Eqs.~\eqref{eq51}--\eqref{eq53} reduce to Eqs.~\eqref{nD_bi}--\eqref{nD_approx_bi}. 
\newline

The proof of Theorem \ref{T10} is given in Appendix C.
\newline

As a final step, to verify our approximation Eq.~\eqref{eq51}, we generate $1000$ directed networks with $N=10000$ for each out-degree combination $(k_1,k_2,\alpha)$. For each network with the same out-degree combination $(k_1,k_2,\alpha)$, we randomly remove a fraction $p$ of links and get the value of $n_D$, and then repeat this process for $1000$ times. Thus, the fraction of driver nodes $n_D$ for a combination $(k_1,k_2,\alpha,p)$ is the average fraction of driver nodes in $10^{6}$ realizations.

Table 4 shows the comparison between Eq.~\eqref{eq51} and simulations. In most cases, the relative errors between Eq.~\eqref{eq51} and simulations are small. We conclude from Table 4 that the simulations are a very robust fit with our approximation Eq.~\eqref{eq51}.
%\begin{figure}[tb]\centering 
%	\includegraphics[width=14cm]{final_figures/bi-modal.PNG}
%	\vspace*{-5mm}
%	\caption{Comparing Eqs.(7) with simulation results.}
%	\label{fig:example}
%\end{figure}  

\begin{table}[ht]
\centering
\caption{Comparison of approximation Eq.~\eqref{eq51} with simulation results.}
\label{tab:outiner_results_simappann_2}
\resizebox{\textwidth}{!}{
\begin{tabular}{l|l|l|l|l|l|l|l|l|l}
\hline
\multirow{2}{*}{$k_1$} &
\multirow{2}{*}{$k_2$} &
\multirow{2}{*}{$k$} &
\multirow{2}{*}{$\alpha$} &
\multicolumn{4}{|c|}{\textbf{Eq.~\eqref{eq51}}} &
\multicolumn{2}{|c}{\textbf{Simulation}} 
\\
\cline{5-10}  %  \cline用于画横线 \cline{i-j}表示从第i列画到第j列

& & & &$p=0.2$ & $r$ & $p=0.5$ & $r$ & $p=0.2$ & $p=0.5$\\

\hline
1  &3 &2.5 &0.25 & 0.251484       &0.50\%       & 0.541569    &0.19\%     &0.252746 &0.540569\\ \hline
1  &3 &2 &0.5  & 0.340662      &0.41\%         & 0.627028       &1.29\%        &0.342065 &0.619028\\ \hline
1  &3 &1.5 &0.75 & 0.431100       &0.29\%       & 0.709013         &2.21\%      &0.432370 &0.693714\\ \hline
2  &4 &3.5 &0.25  & 0.122113        &0.25\%       & 0.410770       &0.29\%        &0.121813 &0.409569\\ \hline
2  &4 &3 &0.5 & 0.183813        &1.32\%       & 0.476848       &0.43\%        &0.186273 &0.474822\\ \hline
2  &4 &2.5 &0.75 & 0.247667   &0.80\%          & 0.535501     &0.43\%          &0.245692 &0.537824\\ \hline
2  &6 &5 &0.25 & 0.033257       &0.87\%        & 0.299464       &0.52\%        &0.033549 &0.297913\\ \hline
2  &6 &4 &0.5 & 0.094631      &1.86\%         & 0.435961      &1.48\%        &0.096426 &0.429607\\ \hline
2  &6 &3 &0.75 & 0.216405       &0.93\%       & 0.514376    &3.06\%     &0.218443 &0.499125\\ \hline
2  &8 &6.5 &0.25  & 0.008650      &0.06\%         & 0.101497       &17.49\%        &0.008655 &0.123010\\ \hline
2  &8 &5 &0.5 & 0.037450        &0.81\%       & 0.406573         &0.15\%      &0.037150 &0.405974\\ \hline
2  &8 &3.5 &0.75  & 0.204397        &0.017\%       & 0.505728       &1.00\%        &0.204363 &0.510815\\ \hline
4  &6 &5.5 &0.25 & 0.020441        &5.68\%       & 0.163736       &0.14\%        &0.021671 &0.163504\\ \hline
4  &6 &5 &0.5 & 0.032167    &4.57\%          & 0.229064     &0.44\%          &0.033706 &0.228061\\ \hline
4  &6 &4.5 &0.75 & 0.050380       &1.00\%        & 0.288043       &0.80\%        &0.049880 &0.285759\\ \hline
4  &8 &7 &0.25 & 0.005504      &1.47\%         & 0.058532      &0.33\%        &0.005586 &0.058338\\ \hline
4  &8 &6 &0.5 & 0.013368       &0.077\%        & 0.135230       &0.42\%        &0.013357 &0.134664\\ \hline
4  &8 &5 &0.75 & 0.033187     &0.27\%         & 0.265665      &0.93\%        &0.033275 &0.263211\\ \hline

\end{tabular}}
\end{table}

\section{Conclusion}

In this paper, we correct the formula given in \cite{Komareji} for the minimum number of driver nodes for a specific class of swarm signaling networks, which are characterised by a regular out-degree. We then generalize the results by considering SSNs with a regular out degree $k$ where a fraction $p$ of the links is unavailable. For this case we derive an implicit equation, whose solution leads to the minimum number of driver nodes. We find that our approximation fits well with simulation results.  
Finally, we relax the condition that the out-degree is regular and look into bi-modal out-degree distributions. For this case we also consider scenarios with unavailable links. We derive an implicit equation and verify its accuracy. We find that our approximation for bi-modal out-degree distribution fits well with simulation results.

\section*{Acknowledgment}

This research was supported by the China Scholarship Council (No. 201706220113).


\section*{Appendix A}
Here we will give the proof of Theorem \ref{T2}.
The out-degree distribution $P_{\text{out}}(\cdot)$ for the unperturbed network is given in Eq.~\eqref{Pout}.
Let us denote the out-degree distribution for the perturbed network by $\bar{P}_{\text{out}}(\cdot)$. Then it follows from Lemma \ref{L1} and Eq.~\eqref{Pout} that
\begin{equation}\label{Pout_bar}
\bar{P}_{\text{out}}(k_{\text{out}})=(1-p)^{k_{\text{out}}}\sum^{N-1}_{j=k_{\text{out}}}\binom{j}{k_{\text{out}}}p^{j-k_{\text{out}}}\delta(k-j).
\end{equation}
Therefore we obtain 
\begin{equation}\label{Pout_bar_large}
    \bar{P}_{\text{out}}(k_{\text{out}})=0, 
\end{equation}
if $k_{\text{out}} > k$ and
\begin{equation}\label{Pout_bar_small}
    \bar{P}_{\text{out}}(k_{\text{out}})=(1-p)^{k_{\text{out}}}\binom{k}{k_{\text{out}}}p^{k-k_{\text{out}}}
\end{equation}
if $k_{\text{out}} \leq k$.
From this we get
\begin{equation}
\begin{split}
    \bar{G}_{\text{out}}(x)=\sum^{\infty}_{k_{\text{out}}=0}\bar{P}_{\text{out}}(k_{\text{out}})x^{k_{\text{out}}}=
    \sum^{k}_{k_{\text{out}}=0}(1-p)^{k_{\text{out}}}\binom{k}{k_{\text{out}}}p^{k-k_{\text{out}}}x^{k_{\text{out}}}=\\
    \sum^{k}_{k_{\text{out}}=0}\binom{k}{k_{\text{out}}}((1-p)x)^{k_{\text{out}}}p^{k-k_{\text{out}}}=(p+(1-p)x)^k.
\end{split}
\end{equation}
This proves that Eq.~\eqref{Gout_bar} holds.
\newline
\newline
We assumed that the in-degree distribution of the original graph follows a Poisson distribution, see \eqref{Pin} but for finite $N$ the actual distribution is binomial. However, for $N \longrightarrow \infty$ the limiting distribution is indeed Poissonian. Therefore, for proving that Eq.~\eqref{Gin_bar} holds, we will use Lemma \ref{L1} with $N=\infty$.
The in-degree distribution $P_{\text{in}}(\cdot)$ for the unperturbed network is given in Eq. \eqref{Pin}.
Let us denote the in-degree distribution for the perturbed network by $\bar{P}_{\text{in}}(\cdot)$. Then it follows from Lemma \ref{L1} and Eq.~\eqref{Pin} that
\begin{equation}\label{Pin_bar}
\bar{P}_{\text{in}}(k_{\text{in}})=(1-p)^{k_{\text{in}}}\sum^{\infty}_{j=k_{\text{in}}}\binom{j}{k_{\text{in}}}p^{j-k_{\text{in}}}\frac{k^{j}}{j!}e^{-k}.
\end{equation}
   
From this we get
\begin{equation}
\begin{split}
    \bar{G}_{\text{in}}(x)=\sum^{\infty}_{k_{\text{in}}=0}\bar{P}_{\text{in}}(k_{\text{in}})x^{k_{\text{in}}}=
    \sum^{\infty}_{k_{\text{in}}=0}(1-p)^{k_{\text{in}}}\sum^{\infty}_{j=k_{\text{in}}}
    \binom{j}{k_{\text{in}}}p^{j-k_{\text{in}}}\frac{k^{j}}{j!}e^{-k}x^{k_{\text{in}}}=\\
    e^{-k}\sum^{\infty}_{k_{\text{in}}=0}(\frac{(1-p)x}{p})^{k_{\text{in}}}\sum^{\infty}_{j=k_{\text{in}}}
    \binom{j}{k_{\text{in}}}\frac{(pk)^{j}}{j!}=\\
    e^{-k}\sum^{\infty}_{k_{\text{in}}=0}(\frac{(1-p)x}{p})^{k_{\text{in}}}\sum^{\infty}_{j=k_{\text{in}}}
    \frac{1}{k_{\text{in}}!}\frac{(pk)^{j}}{(j-k_{\text{in}})!}=\\
     e^{-k}\sum^{\infty}_{k_{\text{in}}=0}(\frac{(1-p)x}{p})^{k_{\text{in}}}\frac{1}{k_{\text{in}}!}\sum^{\infty}_{j=k_{\text{in}}}\frac{(pk)^{j-k_{\text{in}}}(pk)^{k_{\text{in}}}}{(j-k_{\text{in}})!}=\\
     e^{-k}\sum^{\infty}_{k_{\text{in}}=0}(\frac{(1-p)x}{p})^{k_{\text{in}}}\frac{(pk)^{k_{\text{in}}}}{k_{\text{in}}!}\sum^{\infty}_{\hat{j}=0}\frac{(pk)^{\hat{j}}}{\hat{j}!}=\\
     e^{-k}\sum^{\infty}_{k_{\text{in}}=0}\frac{(k(1-p)x)^{k_{\text{in}}}}{k_{\text{in}}!}e^{pk}=\\
     e^{-k}e^{k(1-p)x}e^{pk}=e^{-k(1-p)(1-x)}.
\end{split}
\end{equation}
This proves that Eq.~\eqref{Gin_bar} holds.
\newline
\newline
Next we will prove Theorem \ref{T3}.
Using the same notation as before, it follows from Eq.~\eqref{Hout} that for the perturbed system the generating function $\bar{H}_{\text{out}}(x)$ is given by

\begin{equation}
    \bar{H}_{\text{out}}(x)=\sum^{\infty}_{k_{\text{out}}=1}\frac{k_{\text{out}}\bar{P}_{\text{out}}(k_{\text{out}})}{\langle k_{\text{out}}\rangle}x^{k_{\text{out}}-1}
\end{equation}
Then, using Eqs.~\eqref{Pout_bar_large}--\eqref{Pout_bar_small} we obtain
\begin{equation}
\begin{split}
    \bar{H}_{\text{out}}(x)=\sum^{k}_{k_{\text{out}}=1}\frac{k_{\text{out}}(1-p)^{k_{\text{out}}}\binom{k}{k_{\text{out}}}p^{k-k_{\text{out}}}}{k(1-p)}x^{k_{\text{out}}-1}=\\
    \sum^{k}_{k_{\text{out}}=1}\binom{k-1}{k_{\text{out}}-1}p^{k-k_{\text{out}}}((1-p)x)^{k_{\text{out}}-1}=\\
    \sum^{k-1}_{m=0}\binom{k-1}{m}p^{k-1-m}((1-p)x)^{m}=
    (p+(1-p)x)^{k-1}.
\end{split}    
\end{equation}
Finally, we prove Eq.~\eqref{Hin_bar}.\\
Using the same notation as before, it follows from Eq.~\eqref{Hin} that for the perturbed system the generating function $\bar{H}_{\text{in}}(x)$ is given by

\begin{equation}
    \bar{H}_{\text{in}}(x)=\sum^{\infty}_{k_{\text{in}}=1}\frac{k_{\text{in}}\bar{P}_{\text{in}}(k_{\text{in}})}{\langle k_{\text{in}}\rangle}x^{k_{\text{in}}-1}.
\end{equation}
Then, using Eq.~\eqref{Pin_bar} we obtain
\begin{equation}
\begin{split}
    \bar{H}_{\text{in}}(x)=\sum^{\infty}_{k_{\text{in}}=1}\frac{ k_{\text{in}}(1-p)^{k_{\text{in}}}}{k(1-p)} \sum^{\infty}_{j=k_{\text{in}}}
    \binom{j}{k_{\text{in}}}p^{j-k_{\text{in}}}\frac{k^{j}}{j!}e^{-k}x^{k_{\text{in}}-1}=\\
    e^{-k}\sum^{\infty}_{k_{\text{in}}=1}\frac{k_{\text{in}}(k(1-p)x)^{k_{\text{in}}}}{xk(1-p)k_{\text{in}}!}e^{pk}=e^{-k+pk}\sum^{\infty}_{k_{\text{in}}=1}\frac{(k(1-p)x)^{k_{\text{in}}-1}}{(k_{\text{in}}-1)!}=\\
    e^{-k+pk}\sum^{\infty}_{m=0}\frac{(k(1-p)x)^{m}}{m!}=
    e^{-k+pk+k(1-p)x}=e^{-k(1-p)(1-x)}.
\end{split}    
\end{equation}
This finishes the proof of Theorem \ref{T3}.
\newline

Proof of Theorem \ref{T4}.\\
Using Theorem \ref{T2} and \ref{T3}, the set of equations \eqref{eq1}--\eqref{eq4} becomes

\begin{equation}\label{eq1_bar}
w_1=(p+(1-p)\hat{w}_2)^{k-1}    
\end{equation}
\begin{equation}\label{eq2_bar}
\hat{w}_2=1-e^{-k(1-p)w_1}    
\end{equation}
\begin{equation}\label{eq3_bar}
w_2=1-(p+(1-p)(1-\hat{w}_1))^{k-1}    
\end{equation}
\begin{equation}\label{eq4_bar}
\hat{w}_1=e^{-k(1-p)(1-w_2)}    
\end{equation}
By setting $\hat{w}_2=1-\hat{w}_1$ and $w_1=1-w_2$, it follows that the pair of Eqs.~\eqref{eq1_bar}--\eqref{eq2_bar} is equivalent to the pair of Eqs.~\eqref{eq3_bar}--\eqref{eq4_bar} .\\
From this it follows that $n_D$ in Eq.~\eqref{nD} becomes
\begin{equation}
    n_D=\bar{G}_{\text{out}}(1-\hat{w}_1)+\bar{G}_{\text{in}}(w_2)-1+k(1-p)\hat{w}_1(1-w_2)
\end{equation}
Using Eqs.~\eqref{Gout_bar}, \eqref{Gin_bar} and \eqref{eq4_bar}, this leads to Eq.~\eqref{nD_bar}.
Furthermore, Eq.~\eqref{w2_eq_bar} follows from the substitution of $\hat{w}_1$ given in Eq.~\eqref{eq4_bar} into Eq.~\eqref{eq3_bar}.
\newline
\newline
Finally, we prove that Eq.~\eqref{nD_bar_approx} holds.
First, we rewrite Eq.~\eqref{nD_bar} as
\begin{equation}\label{nD_bar_rewrite}
    n_D=(p+(1-p)(1-\hat{w}_1))^k - 1 + \hat{w}_1 + k(1-p)(1-w_2)\hat{w}_1,
\end{equation}
where $\hat{w}_1$ satisfies
\begin{equation}
    \hat{w}_1 = e^{-k(1-p)(p+(1-p)(1-\hat{w}_1))^{k-1}}.
\end{equation}
Therefore, for large $k$ we obtain
\begin{equation}\label{hat_w1_approx}
    \hat{w}_1 \approx e^{-k(1-p)},
\end{equation}
while from Eq.~\eqref{eq3_bar} we get
\begin{equation}\label{w2_approx}
    1-w_2 = (p+(1-p)(1-\hat{w}_1))^{k-1} \approx 1-(1-p)(k-1)\hat{w}_1.
\end{equation}
Then plugging Eqs.~\eqref{hat_w1_approx} and \eqref{w2_approx} into Eq.~\eqref{nD_bar_rewrite} yields
\begin{equation}
\begin{split}
    n_D\approx 1-(1-p)k\hat{w}_1 - 1 + \hat{w}_1 + k(1-p)(1-(1-p)(k-1)\hat{w}_1)\hat{w}_1=\\
    1-(1-p)k\hat{w}_1 - 1 + \hat{w}_1 + k(1-p)\hat{w}_1  -(1-p)^2k(k-1)\hat{w}_1^2\approx \hat{w}_1\approx e^{-k(1-p)}.
\end{split}
\end{equation}
This completes the proof of Theorem \ref{T4}.

\section*{Appendix B}
Proof of Theorem \ref{T5}.\\
Let us denote the out-degree distribution for the considered network by  $\hat{P}_{\text{out}}(\cdot)$. Then it holds that
\begin{equation}\label{Pout_bi}
\hat{P}_{\text{out}}(k_{\text{out}})=\alpha \delta(k_{\text{out}}-k_1) +(1-\alpha) \delta(k_{\text{out}}-k_2).
\end{equation}
Then, denoting the generating function for the out-degree distribution by $\hat{G}_{\text{out}}$, we get
\begin{equation}
\begin{split}
    \hat{G}_{\text{out}}(x)=\sum^{\infty}_{k_{\text{out}}=0}\hat{P}_{\text{out}}(k_{\text{out}})x^{k_{\text{out}}}=\\
    \sum^{\infty}_{k_{\text{out}}=0}(\alpha \delta(k_{\text{out}}-k_1) +(1-\alpha) \delta(k_{\text{out}}-k_2))x^{k_{\text{out}}}=\alpha x^{k_1} +(1-\alpha) x^{k_2}
    .
\end{split}
\end{equation}

Let us denote the in-degree distribution for the considered network by  $\hat{P}_{\text{in}}(\cdot)$, which for large $N$ will approach a Poisson distribution with average $k=\alpha k_1 + (1-\alpha)k_2$. Then it holds that
\begin{equation}\label{Pin_hat}
\hat{P}_{\text{in}}(k_{\text{in}})=\frac{k^{k_{\text{in}}}}{k_{\text{in}}!}e^{-k}.
\end{equation}
Then, denoting the generating function for the in-degree distribution by $\hat{G}_{\text{in}}$, we get
\begin{equation}
\begin{split}
    \hat{G}_{\text{in}}(x)=\sum^{\infty}_{k_{\text{in}}=0}\hat{P}_{\text{in}}(k_{\text{in}})x^{k_{\text{in}}}=
    \sum^{\infty}_{k_{\text{in}}=0}\frac{k^{k_{\text{in}}}}{k_{\text{in}}!}e^{-k}x^{k_{\text{in}}}=\\
    e^{-k}\sum^{\infty}_{k_{\text{in}}=0}\frac{(kx)^{k_{\text{in}}}}{k_{\text{in}}!}=e^{-k}e^{kx}=e^{-k(1-x)}
    .
\end{split}
\end{equation}
This finishes the proof of Theorem \ref{T5}.
\newline
\newline
Proof of Theorem \ref{T6}.\\

Using the same notation as before, it follows from Eq.~\eqref{Hout} that the generating function $\hat{H}_{\text{out}}(x)$ is given by

\begin{equation}
    \hat{H}_{\text{out}}(x)=\sum^{\infty}_{k_{\text{out}}=1}\frac{k_{\text{out}}\hat{P}_{\text{out}}(k_{\text{out}})}{\langle k_{\text{out}}\rangle}x^{k_{\text{out}}-1}
\end{equation}
Then, using Eqs.~\eqref{Pout_bi} we obtain
\begin{equation}
\begin{split}
    \hat{H}_{\text{out}}(x)=\sum^{\infty}_{k_{\text{out}}=1}\frac{k_{\text{out}} (\alpha \delta(k_{\text{out}}-k_1) +(1-\alpha) \delta(k_{\text{out}}-k_2))}{k}x^{k_{\text{out}}-1}=\\
    \frac{\alpha k_1 x^{k_1-1}+(1-\alpha) k_2 x^{k_2-1}}{k}.
\end{split}    
\end{equation}
Finally, we prove Eq.~\eqref{Hin_bi}.\\
Using the same notation as before, it follows from Eq.~\eqref{Hin} that the generating function $\hat{H}_{\text{in}}(x)$ is given by

\begin{equation}
    \hat{H}_{\text{in}}(x)=\sum^{\infty}_{k_{\text{in}}=1}\frac{k_{\text{in}}\hat{P}_{\text{in}}(k_{\text{in}})}{\langle k_{\text{in}}\rangle}x^{k_{\text{in}}-1}
\end{equation}
Then, using Eq.~\eqref{Pin_hat} we obtain
\begin{equation}
\begin{split}
    \bar{H}_{\text{in}}(x)=\sum^{\infty}_{k_{\text{in}}=1}\frac{k_{\text{in}}k^{k_{\text{in}}}e^{-k}x^{k_{\text{in}}-1}}{kk_{\text{in}}!} =
    e^{-k}\sum^{\infty}_{k_{\text{in}}=1}\frac{k^{k_{\text{in}}-1}x^{k_{\text{in}}-1}}{(k_{\text{in}}-1)!}-\\
    e^{-k}\sum^{\infty}_{i=0}\frac{(kx)^{i}}{i!}=e^{-k}e^{kx}=e^{-k(1-x)}.
\end{split}    
\end{equation}
This finishes the proof of Theorem \ref{T6}
\newline
\newline
Proof of Theorem \ref{T7}.\\
Using Theorems \ref{T5} and \ref{T6}, the set of Eqs.~\eqref{eq1}--\eqref{eq4} becomes

\begin{equation}\label{eq1_bi}
w_1=\frac{\alpha k_1\hat{w}_2^{k_1-1}+(1-\alpha) k_2\hat{w_2}^{k_2-1}}{k}    
\end{equation}
\begin{equation}\label{eq2_bi}
\hat{w}_2=1-e^{-kw_1}    
\end{equation}
\begin{equation}\label{eq3_bi}
w_2=1-\frac{\alpha k_1(1-\hat{w}_1)^{k_1-1}+(1-\alpha)k_2(1-\hat{w}_1)^{k_2-1}}{k}    
\end{equation}
\begin{equation}\label{eq4_bi}
\hat{w}_1=e^{-k(1-w_2)}    
\end{equation}
By setting $\hat{w}_2=1-\hat{w}_1$ and $w_1=1-w_2$, it follows that the pair of Eqs.~\eqref{eq1_bi}--\eqref{eq2_bi} is equivalent to the pair of equations Eqs.~\eqref{eq3_bi}--\eqref{eq4_bi}.\\
From this it follows that $n_D$ in Eq.~\eqref{nD} becomes
\begin{equation}\label{nD_bi_rewrite_2}
    n_D=\hat{G}_{\text{out}}(1-\hat{w}_1)+\hat{G}_{\text{in}}(w_2)-1+k\hat{w}_1(1-w_2)
\end{equation}
Using Eqs.~\eqref{Gout_bi}, \eqref{Gin_bi} and \eqref{eq4_bi}, this leads to Eq.~\eqref{nD_bi}. Furthermore, Eq.~\eqref{w2_eq_bi} follows from the substitution of $\hat{w}_1$ given in Eq.~\eqref{eq4_bi} into Eq.~\eqref{eq3_bi}.
\newline
\newline
Finally, we prove that Eq.~\eqref{nD_approx_bi} holds.
First, we rewrite Eq.~\eqref{nD_bi} as
\begin{equation}\label{nD_bi_rewrite_3}
    n_D=\alpha(1-\hat{w}_1)^{k_1}+(1-\alpha)(1-\hat{w}_1)^{k_2}-1+\hat{w}_1 + k(1-w_2)\hat{w}_1,
\end{equation}
where $\hat{w}_1$ satisfies
\begin{equation}
\begin{split}
    \hat{w}_1 = e^{-(\alpha k_1 (1-\hat{w}_1)^{k_1-1}+(1-\alpha) k_2 (1-\hat{w}_1)^{k_2-1})} \approx\\
    e^{-(\alpha k_1+(1-\alpha) k_2)+(\alpha k_1(k_1-1)+(1-\alpha)k_2(k_2-1))\hat{w}_1}=\\
    e^{k}e^{(\alpha k_1(k_1-1)+(1-\alpha)k_2(k_2-1))\hat{w}_1}
\end{split}
\end{equation}
Therefore, for large $k$ we obtain
\begin{equation}\label{hat_w1_bi_approx}
    \hat{w}_1 \approx e^{-k},
\end{equation}
while from Eq.~\eqref{eq3_bi} we get
\begin{equation}\label{w2_bi_approx}
\begin{split}
    w_2 \approx 1-\frac{\alpha k_1(1-(k_1-1)\hat{w}_1+(1-\alpha) k_2(1-(k_2-1)\hat{w}_1}{k}=\\
    1-\frac{k-(\alpha k_1(k_1-1)+(1-\alpha)k_2(k_2-1))\hat{w}_1}{k}=\\
    \frac{\alpha k_1(k_1-1)+(1-\alpha)k_2(k_2-1)}{k}\hat{w}_1\equiv \sigma \hat{w}_1
    .
\end{split}
\end{equation}
Then plugging Eqs.~\eqref{hat_w1_bi_approx} and \eqref{w2_bi_approx} into Eq.~\eqref{nD_bi_rewrite_3} yields
\begin{equation}
\begin{split}
    n_D\approx \alpha(1-k_1\hat{w}_1)+(1-\alpha)(1-k_2)\hat{w}_1 -1 +\hat{w}_1 +k(1-\sigma \hat{w}_1)\hat{w}_1=\\
    \alpha-\alpha k_1\hat{w}_1 +1 - \alpha -k_2(1-\alpha)\hat{w}_1 - 1 + \hat{w}_1 + k\hat{w}_1  -k\sigma \hat{w}_1^2 \approx \hat{w}_1=e^{-k}.
\end{split}
\end{equation}
This completes the proof of Theorem \ref{T7}.

\section*{Appendix C}

Using Theorems \ref{T8} and \ref{T9}, the set of Eqs.~\eqref{eq1}--\eqref{eq4} becomes

\begin{equation}\label{eq1tilde}
    \omega_1=\frac{\alpha{k_1(p+(1-p){\hat\omega_2})^{k_1-1}}+(1-\alpha){k_2(p+(1-p){\hat\omega_2})^{k_2-1}}}{k}
\end{equation}

\begin{equation}\label{eq2tilde}
    1-\omega_2=\frac{\alpha{k_1(p+(1-p){(1-\hat\omega_1)})^{k_1-1}}+(1-\alpha){k_2(p+(1-p){(1-\hat\omega_1)})^{k_2-1}}}{k}
\end{equation}

\begin{equation}\label{eq3tilde}
    \hat\omega_1=e^{-k(1-p)(1-\omega_2)}
\end{equation}

\begin{equation}\label{eq4tilde}
    1-\hat\omega_2=e^{-k(1-p)\omega_1}
\end{equation}

By setting $\hat\omega_2 =1-\hat\omega_1$ and $\omega_2=1-\omega_1$, it follows that the pair of Eqs.~\eqref{eq2tilde}--\eqref{eq3tilde} is
equivalent to the pair of Eqs.~\eqref{eq1tilde}--\eqref{eq4tilde}. Then by using Eq.~\eqref{nD}, we get
\begin{equation}\label{eqnDtilde}
\begin{split}
    n_D=\alpha{(p+(1-p)(1-e^{-k(1-\omega_2)}))}^{k_1}+(1-\alpha)(p+(1-p)(1-e^{-k(1-\omega_2)}))^{k_2}\\-1+e^{-k(1-p)(1-\omega_2)}+k(1-p)e^{-k(1-\omega_2)}(1-\omega_2)
\end{split}    
\end{equation}
where $w_2$ is the solution of Eqs.~\eqref{eq2tilde}--\eqref{eq3tilde}.
This proves that Eq.~\eqref{eq51} holds.

Finally, we prove that Eq.~\eqref{eq53} holds.
From Eqs.~\eqref{eq2tilde}--\eqref{eq3tilde} it follows that 
\begin{equation}\label{eqw1tilde}
\begin{split}
    \hat\omega_1=e^{-(1-p)(\alpha{k_1(p+(1-p){(1-\hat\omega_1)})^{k_1-1}}+(1-\alpha){k_2(p+(1-p){(1-\hat\omega_1)})^{k_2-1}})} \approx \\
    e^{-k(1-p)}e^{(1-p)^2(\alpha k_1(k_1-1)+(1-\alpha)k_2(k_2-1))\hat{w}_1}
\end{split}
\end{equation}
Therefore, for large $k$ we obtain
\begin{equation}\label{hat_w1_bi_approx2}
    \hat{w}_1 \approx e^{-k(1-p)},
\end{equation}
Similarly, from Eq.~\eqref{eq2tilde} we can deduce
\begin{equation}\label{bijna}
    w2 \approx \frac{(1-p)(\alpha k_1(k_1-1)+(1-\alpha)k_2(k_2-1))}{k}\hat{w}_1\equiv \sigma \hat{w}_1
\end{equation}

Substitution of Eq.~\eqref{hat_w1_bi_approx2} and Eq.~\eqref{bijna} into Eq.~\eqref{eqnDtilde}, we obtain

\begin{equation}
\begin{split}
    n_D\approx e^{-\bar{k}(1-p)}
\end{split}
\end{equation}
This completes the proof of Theorem \ref{T10}.

%\bibliographystyle{comnet}
%\bibliography{sample}

%
% once the .bbl file has been generated then place the text in your article.

% To get the unnumbered reference style the author should use [unnumbib]
%as an option in the document class.  For example: \documentclass[unnumbib]{comnet}

% \begin{thebibliography}{99}

% \bibitem{Rottmann:2010a}
% {\sc Rottmann-Matthes, J.} (2011a) Linear stability of traveling
% waves in nonstrictly hyperbolic PDES.\break {\em J. Dynam.
% Differential Equations}, \textbf{23}, 365--393.

% \bibitem{Rottmann:2011a}
% {\sc Rottmann-Matthes, J.} (2011b) Stability and freezing of
% nonlinear waves in first-order hyperbolic PDEs. Preprint
% 11-016, CRC 701, Bielefeld University.

% \bibitem{Rottmann:2011b}
% {\sc Rottmann-Matthes, J.} (2012) Stability of
% parabolic-hyperbolic traveling waves. Preprint 12-005, CRC
% 701, Bielefeld University.

% \bibitem{RowleyKevrekidisMarsdenLust:2003}
% {\sc Rowley, C. W., Kevrekidis, I. G., Marsden, J. E. \& Lust,
% K.} (2003) Reduction and reconstruction for self-similar
% dynamical systems. {\em Nonlinearity}, \textbf{16},
% 1257--1275.

% \end{thebibliography}

\end{document}